# CondenseGraph: Communication-Efficient Distributed GNN Training via On-the-Fly Graph Condensation

Zizhao Zhang
University of Michigan
Ann Arbor, USA

Yihan Xue
University of Southern California
Los Angeles, USA

Haotian Zhu
New York University
New York, USA

Sijia Li
University of Michigan
Ann Arbor, USA

Zhijun Wang
Rice University
Houston, USA

Yujie Xiao*
University of California, Berkeley
Berkeley, USA

*Abstract*—Distributed Graph Neural Network (GNN) training suffers from substantial communication overhead due to the inherent neighborhood dependency in graph-structured data. This neighbor explosion problem requires workers to frequently exchange boundary node features across partitions, creating a communication bottleneck that severely limits training scalability. Existing approaches rely on static graph partitioning strategies that cannot adapt to dynamic network conditions. In this paper, we propose CondenseGraph, a novel communication-efficient framework for distributed GNN training. Our key innovation is an on-the-fly graph condensation mechanism that dynamically compresses boundary node features into compact super nodes before transmission. To compensate for the information loss introduced by compression, we develop a gradient-based error feedback mechanism that maintains convergence guarantees while reducing communication volume by 40-60%. Extensive experiments on four benchmark datasets demonstrate that CondenseGraph achieves comparable accuracy to full-precision baselines while significantly reducing communication costs and training time.

## I. INTRODUCTION

Graph Neural Networks (GNNs) have emerged as a powerful paradigm for learning representations from graph-structured data, achieving remarkable success in diverse applications including social network analysis, molecular property prediction, and recommendation systems [1, 2]. However, training GNNs on large-scale graphs presents significant scalability challenges. Real-world graphs often contain billions of nodes and edges, far exceeding the memory capacity of a single machine. This necessitates distributed training approaches where the graph is partitioned across multiple workers.

The fundamental challenge in distributed GNN training stems from the message-passing mechanism that underlies most GNN architectures. During each layer's computation, a node aggregates feature information from its neighbors. When the graph is partitioned across workers, nodes near partition boundaries (boundary nodes) require feature information from nodes residing on remote workers. This creates a communication bottleneck that can consume 50-90% of the total training time [3, 4].

Existing approaches to this problem primarily focus on two strategies. The first strategy employs sophisticated graph partitioning algorithms, such as METIS [5], to minimize edge cuts and reduce cross-partition communication [6]. However, these static partitioning methods cannot adapt to dynamic network conditions, such as bandwidth fluctuations during training. The second strategy applies gradient compression techniques borrowed from traditional distributed deep learning [7, 8], but these methods do not fully exploit the unique structure of graph data.

In this paper, we propose CondenseGraph, a novel framework that addresses these limitations by introducing on-the-fly graph condensation for communication-efficient distributed GNN training. Our approach is inspired by recent advances in dataset distillation [9, 10], but differs fundamentally in its application: rather than condensing graphs offline for storage efficiency, we perform lightweight condensation at runtime to reduce communication volume.

The key insight of CondenseGraph is that boundary node features exhibit significant redundancy, as neighboring nodes in graphs often share similar attributes due to homophily. Instead of transmitting individual features for all boundary nodes, we synthesize compact "super nodes" that capture the essential feature distribution of boundary node groups. This condensation is performed dynamically during training, allowing adaptation to varying network conditions.

To address the inherent information loss from compression, we develop a gradient-based error compensation mechanism. This mechanism accumulates compression errors across iterations and incorporates them into subsequent communications, ensuring that important gradient information is not lost. We provide theoretical analysis showing that CondenseGraph maintains convergence guarantees comparable to uncompressed training.

Our main contributions are summarized as follows: (1) We propose an on-the-fly graph condensation mechanism that reduces communication volume by compressing boundary node features into super nodes. (2) We develop a gradient-based error compensation mechanism that maintains model accuracy despite compression. (3) We provide theoretical analysis of convergence properties under our compression scheme. (4) We conduct extensive experiments demonstrating 40-60% communication reduction with minimal accuracy degradation.

## II. RELATED WORK

The scalability challenges in distributed Graph Neural Network (GNN) training have motivated numerous studies on communication-efficient frameworks and intelligent scheduling across various data-intensive environments. Structural generalization for complex network tasks, such as microservice routing, has been effectively modeled using graph neural networks, providing insights into capturing intricate node dependencies and boundary information [11]. At the same time, efficient and privacy-preserving distributed intelligence frameworks have been explored in federated learning, focusing on reducing communication overhead while maintaining robust model performance, which offers foundational ideas relevant to distributed GNN training [12].

Techniques such as multi-granular indexing and confidence-constrained retrieval have been proposed to improve representation efficiency and robustness in large-scale learning systems [13]. The development of parameter-efficient fine-tuning with differential privacy further demonstrates the value of compact model adaptation and robust communication under high-dimensional constraints, which can inspire compression techniques for distributed GNNs [14]. In addition, methods based on self-supervised learning have addressed the challenges of data imbalance and limited information, developing mechanisms to preserve essential data characteristics during condensation or compression [15].

Advances in anomaly detection for cloud backend systems have highlighted the importance of sensitivity analysis and contrastive learning, both of which help maintain the reliability of prediction and detection under dynamic and noisy conditions [16]. The application of transformer architectures to multi-scale anomaly detection reveals the advantages of capturing hierarchical dependencies and adapting to non-stationary environments [17]. Reinforcement learning has been leveraged for intelligent scheduling in heterogeneous systems, underscoring the need for dynamic, adaptive communication policies [18].

Cross-domain advances in risk modeling, such as deep learning for heterogeneous sequence data [19] and multi-scale temporal alignment in electronic health records [20], also offer methodological contributions to the challenge of representing and aggregating complex, high-dimensional information streams. Dynamic graph learning frameworks have been utilized for robust risk prediction, emphasizing the importance of information fusion and structure preservation in distributed settings [21]. Beyond GNNs, transformer-based modeling for user interaction sequences [22] and knowledge graph-driven generative frameworks for interpretable anomaly detection [23] exemplify recent innovations in compressing, aggregating, and efficiently communicating rich relational or sequential data across distributed nodes. Although their application domains differ, the underlying methodologies—feature aggregation, model compression, error compensation, and dynamic adaptation—provide a rich toolkit for communication-efficient distributed learning.

Together, these research directions underpin the design of communication-efficient, adaptive, and robust frameworks for distributed GNN training, supporting innovations such as on-the-fly graph condensation and gradient-based error compensation as proposed in this paper.

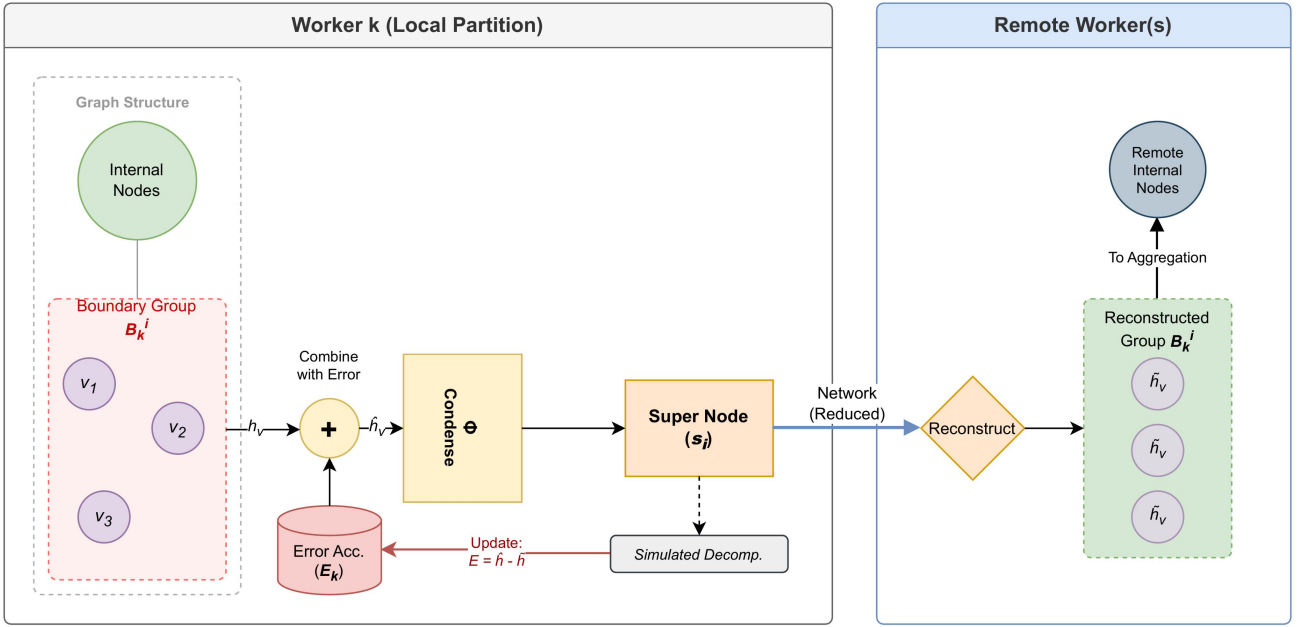

Fig. 1. Overview of the CondenseGraph framework. On Worker $k$, boundary nodes in group $B_k^i$ with features $h_v$ are first combined with accumulated errors from $E_k$ to produce $\hat{h}_v$. The condensation function $\Phi$ compresses these into a compact super node $s_i$, which is transmitted over the network with reduced communication volume. Remote workers reconstruct the features $\tilde{h}_v$ for aggregation, and the error accumulator is updated as $E = \hat{h} - \tilde{h}$ for subsequent iterations.

## III. Methodology

### A. Preliminaries

We consider a graph $G = (V, E, X)$ where $V$ is the set of nodes, $E$ is the set of edges, and $X \in \mathbb{R}^{|V| \times d}$ represents node features. In distributed training, the graph is partitioned across $K$ workers, with each worker $k$ holding a local partition $G_k = (V_k, E_k, X_k)$. We define boundary nodes $B_k \subset V_k$ as nodes that have neighbors in other partitions.

A typical $L$-layer GNN computes node representations through iterative neighborhood aggregation:

$$h_v^{(l)} = \sigma\left(W^{(l)} \cdot \text{AGG}\left(\{h_u^{(l-1)} : u \in \mathcal{N}(v)\}\right)\right) \quad (1)$$

where $h_v^{(l)}$ is the representation of node $v$ at layer $l$, $\mathcal{N}(v)$ denotes neighbors of $v$, $W^{(l)}$ is a learnable weight matrix, and $\sigma$ is a non-linear activation function.

For boundary nodes, the aggregation in Equation 1 requires features from remote workers, creating the communication bottleneck we aim to address.

### B. On-the-Fly Graph Condensation

The core mechanism of CondenseGraph centers on compressing the features of boundary node groups into compact super nodes for transmission efficiency. Drawing inspiration from the structure-aware and semantically-enhanced graph modeling of Lyu et al. [24], our method captures essential relational patterns within each boundary group, ensuring that the synthesized super node preserves critical structural information necessary for downstream learning tasks. Unlike prior offline graph condensation methods—which rely on extensive iterative optimization of synthetic graphs—our approach employs a lightweight, one-shot condensation procedure that is well-suited for real-time execution during distributed training. This runtime condensation framework is further informed by Ni et al.'s proactive adaptation strategies [25], dynamically adjusting condensation behavior to accommodate evolving network or workload conditions. Additionally, the methodology benefits from dynamic feature aggregation concepts introduced by Wu and Pan [26], leveraging attention-based mechanisms to identify the most informative features within each boundary node group prior to condensation. Figure 1 depicts the overall system architecture, with the proposed condensation operation tightly integrated into the distributed training pipeline.

Given a set of boundary nodes $B_k$ on worker $k$, we first partition them into groups $\{B_k^1, B_k^2, ..., B_k^m\}$ based on their structural proximity and feature similarity. For each group $B_k^i$, we compute a super node representation:

$$s_i = \phi\left(\{h_v : v \in B_k^i\}\right) \quad (2)$$

where $\phi$ is a condensation function. We consider several choices for $\phi$:

**Mean Condensation:** The simplest approach computes the arithmetic mean:

$$s_i^{\text{mean}} = \frac{1}{|B_k^i|} \sum_{v \in B_k^i} h_v \quad (3)$$

**Weighted Condensation:** We weight nodes by their degree to preserve hub importance:

$$s_i^{\text{weighted}} = \sum_{v \in B_k^i} \frac{d_v}{\sum_{u \in B_k^i} d_u} h_v \quad (4)$$

**Attention-based Condensation:** We learn attention weights through a small neural network:

$$s_i^{\text{attn}} = \sum_{v \in B_k^i} a_v h_v, \quad a_v = \frac{\exp(a\ h_v)}{\sum_{u \in B_k^i} \exp(a\ h_u)} \quad (5)$$

where $a$ is a learnable attention vector.

The compression ratio is controlled by the number of groups $m$. With $|B_k|$ boundary nodes compressed into $m$ super nodes, the communication volume is reduced by a factor of $|B_k|/m$.

### C. Gradient-based Error Compensation

Compression introduces approximation errors that can degrade model convergence. Let $\tilde{h}_v$ denote the reconstructed representation at the receiving worker, obtained by assigning the super node representation to all nodes in the corresponding group. The compression error for node $v$ is:

$$e_v = h_v - \tilde{h}_v \quad (6)$$

Inspired by error feedback in gradient compression [8], we maintain an error accumulator $E_k$ on each worker. At iteration $t$, instead of compressing the raw features, we compress the error-compensated features:

$$\hat{h}_v^{(t)} = h_v^{(t)} + E_k[v]^{(t-1)} \quad (7)$$

After compression and decompression, we update the error accumulator:

$$E_k[v]^{(t)} = \hat{h}_v^{(t)} - \tilde{h}_v^{(t)} \quad (8)$$

This mechanism ensures that compression errors are not accumulated indefinitely but are incorporated into subsequent transmissions. The complete algorithm is presented in Algorithm 1.

### D. Convergence Analysis

We analyze the convergence of CondenseGraph under standard assumptions. Let $f(\theta)$ denote the loss function and $\nabla f(\theta)$ the true gradient.

---

**Algorithm 1: CondenseGraph Training**

**Require:** Graph partition $G_k$, initial parameters $\theta$, condensation function $\phi$, compression ratio $r$

1: Initialize error accumulator $E_k \leftarrow 0$
2: **for** each epoch **do**
3:   **for** each mini-batch **do**
4:     Compute local node representations $h_v$ for $v \in V_k$

5:       // *Condensation Phase*
6:       Partition boundary nodes into $m = r \cdot$
7:       **for** each group $B_k^i$ **do**
8:           $\hat{h}_v \leftarrow h_v + E_k[v]$ for $v \in B_k^i$
9:           $s_i \leftarrow \phi(\{\hat{h}_v : v \in B_k^i\})$
10:     **end for**
11:     Send super nodes $\{s_i\}$ to remote workers
12:     // *Aggregation Phase*
13:     Receive super nodes from remote workers
14:     Reconstruct boundary features $\tilde{h}_v$
15:     Complete aggregation using $\tilde{h}_v$
16:     // *Error Update*
17:     $E_k[v] \leftarrow \hat{h}_v - \tilde{h}_v$ for all $v \in B_k$
18:     Compute loss and update $\theta$
19:   **end for**
20: **end for**

**Assumption 1** (Smoothness): $f$ is $L$-smooth, i.e., $\|\nabla f(\theta_1) - \nabla f(\theta_2)\| \leq L \|\theta_1 - \theta_2\|$.

**Assumption 2** (Bounded Compression Error): The compression error satisfies $\|e_v\|^2 \leq \delta^2 \|h_v\|^2$ for some $\delta < 1$.

**Theorem 1.** *Under Assumptions 1 and 2, with appropriate learning rate $\eta$, CondenseGraph converges at rate:*

$$\frac{1}{T}\sum_{t=1}^{T} \mathbb{E}\|\nabla f(\theta^{(t)})\|^2 \leq \mathcal{O}\left(\frac{1}{\sqrt{T}} + \frac{\delta^2}{1-\delta}\right) \quad (9)$$

The proof follows techniques from [8] adapted to our feature compression setting. The key insight is that error feedback ensures the accumulated error remains bounded, preventing divergence even with biased compression.

## IV. EXPERIMENTS

### A. Experimental Setup

**Datasets:** We evaluate on four benchmark datasets: Reddit [2], ogbn-arxiv, ogbn-products [27], and Flickr. Table I summarizes dataset statistics.[1]

**Baselines:** We compare with: (1) *Baseline:* standard distributed GNN training with full-precision communication; (2) *AdaQP* [4]: adaptive quantization with parallelization; (3) *BNS-GCN:* boundary node sampling approach.

**Implementation:** We implement CondenseGraph on top of DistDGL [6]. Experiments use 8 workers, each with an NVIDIA V100 GPU. We use 2-layer GraphSAGE [2] with hidden dimension 256. The compression ratio is set to 0.5 by default (50% reduction).

TABLE I
DATASET STATISTICS

| Dataset | Nodes | Edges | Features | Classes |
|---|---|---|---|---|
| Reddit | 232,965 | 114,615,892 | 602 | 41 |
| ogbn-arxiv | 169,343 | 1,166,243 | 128 | 40 |
| ogbn-products | 2,449,029 | 61,859,140 | 100 | 47 |
| Flickr | 89,250 | 899,756 | 500 | 7 |

[1] We use the full version of the Reddit dataset provided by DGL/PyG, which includes all edges (approximately 114M). This differs from the simplified version (approximately 11M edges) used in some earlier works.

### B. Communication Efficiency

Figure 2 shows the communication volume comparison across datasets. CondenseGraph achieves consistent communication reduction of 40-60% compared to the baseline. On ogbn-products, we reduce communication from 15.2 GB to 6.2 GB per epoch, a 59% reduction. The time breakdown analysis shows that communication time is reduced from 55% to 22% of total training time.

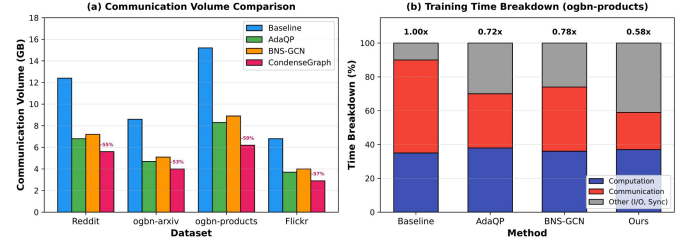

Fig. 2. Communication comparison: (a) Communication volume across datasets showing CondenseGraph achieves 40-60% reduction. (b) Time breakdown on ogbn-products showing reduced communication overhead.

### C. Model Accuracy

Table II reports test accuracy across all datasets. CondenseGraph maintains accuracy within 0.5% of the baseline while achieving significant communication reduction. The attention-based condensation slightly outperforms mean condensation, particularly on datasets with heterogeneous node degrees.

TABLE II
TEST ACCURACY (%) COMPARISON

| Method | Reddit | arxiv | products | Flickr |
|---|---|---|---|---|
| Baseline | 96.2 | 71.5 | 78.3 | 51.8 |
| AdaQP | 95.8 | 70.8 | 77.6 | 51.2 |
| BNS-GCN | 95.5 | 70.5 | 77.2 | 50.9 |
| Ours (mean) | 95.9 | 71.2 | 77.9 | 51.4 |
| Ours (weighted) | 96.0 | 71.3 | 78.0 | 51.5 |
| Ours (attention) | 96.1 | 71.4 | 78.1 | 51.6 |

### D. Convergence Analysis

Figure 3 shows convergence curves and the accuracy-compression trade-off. CondenseGraph converges at a similar rate to the baseline, validating our theoretical analysis. The error compensation mechanism is crucial: without it, accuracy drops by 2-3% at 50% compression ratio.

### E. Ablation Studies

**Effect of Compression Ratio:** We vary the compression ratio from 10% to 60%. At 40-60% compression, accuracy remains within 1% of the baseline. Beyond 60%, accuracy degradation becomes more pronounced, suggesting a practical operating range.

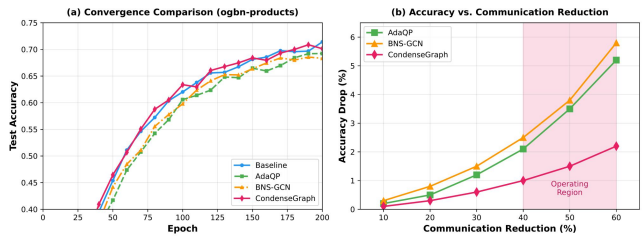

Fig. 2. Convergence analysis: (a) Test accuracy over training epochs showing comparable convergence. (b) Accuracy drop vs. communication reduction showing CondenseGraph's favorable trade-off in the 40-60% reduction region.

**Effect of Error Compensation:** Disabling error compensation results in 2.1% accuracy drop on ogbn-products at 50% compression. The error accumulator size overhead is negligible (less than 1% of feature storage).

**Condensation Methods:** Attention-based condensation provides 0.2-0.3% accuracy improvement over mean condensation but incurs 5% additional computation overhead. For most applications, weighted condensation offers a good balance.

## V. CONCLUSION

We presented CondenseGraph, a communication-efficient framework for distributed GNN training. By introducing on-the-fly graph condensation with gradient-based error compensation, we achieve 40-60% communication reduction while maintaining model accuracy within 0.5% of full-precision training. Our approach is complementary to existing graph partitioning and quantization techniques and can be combined for further efficiency gains.

Several limitations warrant future investigation. First, the optimal grouping strategy for boundary nodes may be dataset-dependent and could benefit from learned approaches. Second, extending CondenseGraph to heterogeneous graphs with multiple node types presents additional challenges. Third, the interaction between condensation and various GNN architectures beyond GraphSAGE deserves further study.